\documentclass[prl,twocolumn,showpacs,superscriptaddress,preprintnumbers,amssymb]{revtex4}
\usepackage{graphicx}

\newcommand{\beq}{\begin{equation}}
\newcommand{\eeq}{\end{equation}}
\newcommand{\beqn}{\begin{eqnarray}}
\newcommand{\eeqn}{\end{eqnarray}}

\begin{document}
\title{Topological Quantum Liquids with Quaternion Non-Abelian Statistics}
\author{Cenke Xu}
\affiliation{Department of Physics, University of California,
Santa Barbara, CA 93106}
\author{Andreas W. W. Ludwig}
\affiliation{Department of Physics, University of California,
Santa Barbara, CA 93106}

\date{\today}

\begin{abstract}

Noncollinear magnetic order is typically characterized by a
``tetrad" ground state manifold (GSM) of three perpendicular
vectors or nematic-directors. We study three types of tetrad
orders in two spatial dimensions, whose GSMs are $\mathrm{SO(3)} =
S^3/Z_2$, $S^3/Z_4$, and $S^3/Q_8$, respectively. $Q_8$ denotes
the non-Abelian quaternion group with eight elements. We
demonstrate that after quantum disordering these three types of
tetrad orders, the systems enter fully gapped liquid phases
described by $Z_2$, $Z_4$, and non-Abelian quaternion gauge field
theories, respectively. The latter case realizes Kitaev's
non-Abelian toric code in terms of a rather simple spin-1 $SU(2)$
quantum magnet. This non-Abelian topological phase possesses a
22-fold ground state degeneracy on the torus arising from the 22
representations of the Drinfeld double of $Q_8$.

\end{abstract}
 \maketitle

The search for quantum liquid states has been one of the main
goals of condensed matter theory for decades. There are in general
two different routes towards this goal, starting from two opposite
limits. The first route is to start with the quantum limit, say
the large-$N$ limit of the SU($N$) antiferromagnet, and approach
the physical system through an $1/N$ expansion. For instance, the
$1/N$ expansion within the slave fermion formalism for the SU($N$)
antiferromagnet leads to the valence bond solid state with no
classical counterpart \cite{sachdev1989}. In our current work, we
will take a second route towards the liquid state, which is by
quantum disordering the semiclassical state. For instance, it is
understood that the valence bond solid state (VBS) naturally
emerges if quantum fluctuations destroy the semiclassical N\'{e}el
order of a spin-1/2 system. This result is based on the
observation that the Skyrmion of the N\'{e}el order of a spin-1/2
antiferromagnet always carries lattice momentum
\cite{haldanemonopole,sachdev1990}.

\begin{figure}
\includegraphics[width=3.0 in]{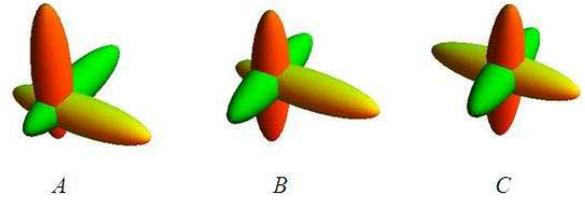}
\caption{Three types of tetrad spin order. Type $A$ phase has
ground state manifold SO(3), which is equivalent to the
configurations of three perpendicular vectors. Type $B$ phase has
GSM $S^3/Z_4$, two of the three perpendicular vectors are headless
directors. In type $C$ phase all three vectors are headless
directors. } \label{nematicfig}
\end{figure}

In general, the ground state manifold (GSM) of spin states can be
written as \beqn \mathrm{GSM} = \mathrm{SU(2)} / \ G, \eeqn where
$G$ represents the unbroken subgroup of the SU(2) spin symmetry in
the ordered phase. $G$ is at least $Z_2$ for spin-1/2 systems,
because physical order parameters should be invariant under spin
rotation by $2\pi$. In the present paper, we will discuss the
quantum disordered phases adjacent to semiclassical spin states
whose unbroken symmetry $G$ is a discrete subgroup of SU(2),
either Abelian or non-Abelian. All these states are
``tetrad-like'' states $i.e.$ the GSM can be represented by three
perpendicular vectors or nematic-directors
(Fig.~\ref{nematicfig}). We will demonstrate that an exotic
non-Abelian topological liquid state can emerge after disordering
a tetrad nematic order of a fairly simple spin-1 system.
Non-Abelian statistics is a much sought-out phenomenon much
discussed in particular in fractional quantum Hall systems
\cite{mooreread}, and more recently also in certain topological
insulators (superconductors) \cite{fu2008}. In the sequel, we will
discuss in turn three tetrad states, Type A, B, and C.

\noindent{\it -- Type A, with G = $Z_2$:} Let us first take $G =
Z_2$, and thus the GSM is now $\mathrm{SU(2)}/Z_2 =
\mathrm{SO(3)}$.
SO(3) is precisely the tetrad manifold, which corresponds to all
the configurations of three perpendicular vectors
(Fig.~\ref{nematicfig}$A$). One example of this case is the well
understood noncollinear spin density wave (SDW), for which the
three perpendicular vectors $\vec{N}_1$, $\vec{N}_2$ and
$\vec{N}_3$ that characterize the GSM are defined as $
\vec{S}(\vec{r}) = \vec{N}_2 \cos(2\vec{Q}\cdot \vec{r}) +
\vec{N}_3 \sin(2\vec{Q}\cdot \vec{r})$; $\vec{N}_1 = \vec{N}_2
\times \vec{N}_3$. Here $\vec{Q}$ is the spiral wave vector of the
SDW. It was pointed out in Ref.~\cite{senthil1994} that if quantum
fluctuations destroy the noncollinear spin density wave (SDW),
one interesting possibility is that the system enters a $Z_2$
liquid state.
On the torus this $Z_2$ liquid ground state has a four-fold
topological degeneracy \cite{kitaev1997}. Let us briefly review
why this is the case.
The most convenient way of parametrizing the manifold SO(3) is by
introducing CP${}^1$ spinor fields $z = (z_1, z_2)^t$ as follows:
\beqn \vec{N}_1 \sim z^\dagger \vec{\sigma} z, \ \vec{N}_2 \sim
\mathrm{Re}[z^t i \sigma^x \vec{\sigma} z], \ \vec{N}_3 \sim
\mathrm{Im}[z^t i \sigma^y \vec{\sigma} z]. \eeqn It is
straightforward to show that the vectors $\vec{N}_a$ are
automatically perpendicular to each other after introducing the
spinor $z_\alpha$. Since all the physical vectors $\vec{N}_a$ are
bilinears of $z_\alpha$, the spinor $z_\alpha$ is effectively
coupled to a $Z_2$ gauge field, which makes $z_\alpha$ equivalent
to $-z_\alpha$.

\begin{figure}
\includegraphics[width=2.8 in]{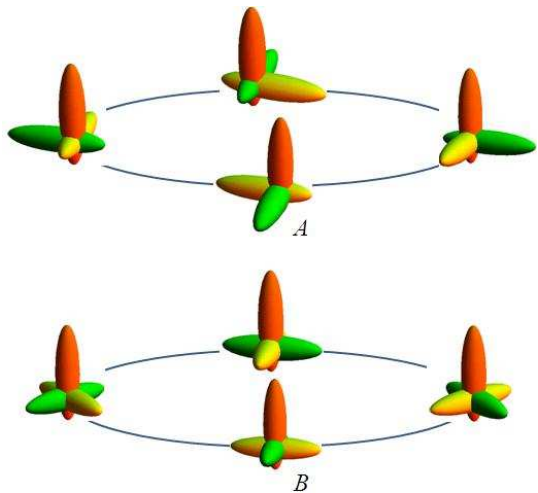}
\caption{The configuration of vison defect in tetrad order A and
half-vison defect in tetrad order B.} \label{defect}
\end{figure}

Since the homotopy group $\pi_1[\mathrm{SO(3)}] = Z_2$, the GSM
SO(3) supports vortex like topological defects with a $Z_2$
conservation law. This type of topological defect is often called
a vison. Pictorially, the vison can be viewed as a configuration
in which (for instance) $\vec{N}_1$ is uniform in space, while
$\vec{N}_2$ and $\vec{N}_3$ have a vortex (Fig.~\ref{defect}$A$).
After we destroy the ordered state with quantum fluctuations, the
spinor $z_\alpha$ is gapped, but the $Z_2$ conservation law of the
vison still persists. This implies that the disordered phase of
the noncollinear SDW is equivalent to the deconfined phase of
$Z_2$ gauge theory, where visons also have a $Z_2$ conservation
law \cite{senthil1994,kitaev1997}. In this phase the gapped spinor
$z_\alpha$ and the vison have mutual semionic statistics, $i.e.$
the wave function picks up a minus sign when $z_\alpha$ encircles
the vison adiabatically.

As was discussed in Ref.~\cite{senthil1994}, the transition
between the ordered phase with GSM SO(3) and the $Z_2$ deconfined
liquid phase is contiuous and belongs to the 3D O(4) universality
class. This is because the bosonic spinor field $z_\alpha$ can
also be viewed as a four component real vector, whose
order-disorder phase transition belongs to the 3D O(4)
universality class. The gapped $Z_2$ gauge field does not
introduce singular corrections in the infrared, $i.e.$ the 3D O(4)
universality class is unaffected by the the presence of the $Z_2$
gauge field \cite{senthil1994}.
Because the physical order parameters are bilinears of the spinor
$z_\alpha$, they acquire a relatively large anomalous dimension as
compared to the standard order parameters at the  Wilson-Fisher
fixed point of the $O(4)$ Heisenberg model. Specifically, within a
five-loop epsilon expansion in $d=4-\epsilon$ dimensions the
scaling dimension of these composite bilinears would be estimated
to be $ \eta_{\vec{N}_a} \approx 1.37$ in $(2+1)$ dimensions
\cite{vicari2002,vicari2003} .

\noindent{\it -- Type B, with G = $Z_4$:} Now let us move to the
type-$B$ tetrad phase. The GSM can be characterized by one vector
and two directors, where again all three vector/directors are
perpendicular to each other (Fig.~\ref{nematicfig}$B$). It is more
convenient to describe this manifold using the following slightly
different representation of $\vec{N}_a$: \beqn && \vec{N}_1 \sim
\mathrm{tr}[\mathcal{Z}^\dagger \vec{\sigma} \mathcal{Z}
\sigma^z], \vec{N}_2 \sim \mathrm{tr}[\mathcal{Z}^\dagger
\vec{\sigma} \mathcal{Z} \sigma^x], \vec{N}_3 \sim
\mathrm{tr}[\mathcal{Z}^\dagger \vec{\sigma} \mathcal{Z}
\sigma^y], \cr\cr && \mathcal{Z} = \phi_0 1 + i \phi_1 \sigma^x +
i \phi_2 \sigma^y + i \phi_3 \sigma^z, \cr\cr && z = (z_1, \
z_2)^t = (\phi_0 + i\phi_3, \ -\phi_2 + i \phi_1)^t. \label{rotor}
\eeqn $\mathcal{Z}$ is a SU(2) matrix, sometimes called the SU(2)
slave rotor field\footnote{These notations were first introduced
in this context in Ref. \cite{Hermele2007}.}. $\mathcal{Z}$ has an
action by $\mathrm{SU(2)_{left}}$ (left multiplication) and by
$\mathrm{SU(2)_{right}}$ (right multiplication). While
$\mathrm{SU(2)_{left}}$ transformations correspond to the physical
SU(2) spin rotation symmetry, $\mathrm{SU(2)_{right}}$
transformations contain the gauge symmetry as a subgroup.

Now let us take $\vec{N}_1$ a vector, while $\vec{N}_2$ and
$\vec{N}_3$ are both headless directors. In order to make
$\vec{N}_2$ and $\vec{N}_3$ headless, we can couple $\mathcal{Z}$
to a gauge field taking values in a group with group elements:
\beqn Z_4 = \{ 1, \ i\sigma^z, \ - 1, \ -i\sigma^z\}.
\label{z4group}\eeqn The gauge field always acts on $\mathcal{Z}$
by right multiplication. Under the gauge transformation \beqn
\mathcal{Z} \rightarrow \mathcal{Z}(\pm i\sigma^z),
\label{gaugetransform}\eeqn both $\vec{N}_2$ and $\vec{N}_3$
reverse direction, while $\vec{N}_1$ remains invariant. Therefore
the type $B$ tetrad phase can be understood as the condensate of
the  SU(2) rotor field $\mathcal{Z}$ (or spinor $z_\alpha$) when
it is coupled to the $Z_4$ gauge field with gauge group $Z_4$ from
Eq.~\ref{z4group}. Unlike the type $A$ case, $\vec{N}_2$ and
$\vec{N}_3$ are no longer themselves physical order parameters due
to the presence of the $Z_4$ gauge field; rather, the physical
order parameter $Q^{ab}_i = N^a_iN^b_i -
\frac{1}{3}(\vec{N}_i)^2$, $i = 2,3$ is of quadrupolar type.

In addition to the vison defect discussed in the type $A$ phase,
the type $B$ phase also has a ``half-vison" defect, $i.e.$ the
configuration in which $\vec{N}_1$ is uniform in space, while
$\vec{N}_2$ and $\vec{N}_3$ have a half vortex
(Fig.~\ref{defect}$B$). This defect has a logarithmically
divergent instead of a confining energy because $\vec{N}_2$ and
$\vec{N}_3$ are nematic directors. We can also describe this half
vortex as a $Z_4$ gauge flux $\pm i\sigma^z$ in the condensate of
$\mathcal{Z}$. After encircling this flux, $\mathcal{Z}$ undergoes
a  gauge transformation as in Eq.~\ref{gaugetransform}, and
$\vec{N}_2$ and $\vec{N}_3$ reverse their directions.


When the rotor field $\mathcal{Z}$ that couples to the $Z_4$ gauge
field
is gapped out, the system is described by a pure $Z_4$-gauge
theory --  a `$Z_4$-liquid' phase. If the system in this phase is
defined on the torus, then there can be four different fluxes
through each cycle of the torus: $0$, $\pi/2$, $\pi$, $3\pi/2$.
Each of these different flux combinations corresponds to an
independent topological sector. There is thus a 16 fold
topological degeneracy on the torus. Recently, one of the authors
of the present paper proposed that the type $B$ phase is an
intermediate phase of the Hubbard model on the honeycomb lattice
\cite{xu2010}, sandwiched between a fully gapped spin liquid phase
and a pure N\'{e}el order with ground state manifold $S^2$. In
Ref.~\cite{xu2010}, the vector $\vec{N}_1$ is the N\'{e}el order,
whereas the directors $\vec{N}_2$ and $\vec{N}_3$ are spin nematic
orders. The transition between type $B$ tetrad order and the $Z_4$
liquid phase also belongs to the 3D O(4) universality class, for
the same reason as in the type $A$ case.


\begin{figure}
\includegraphics[width=2.8 in]{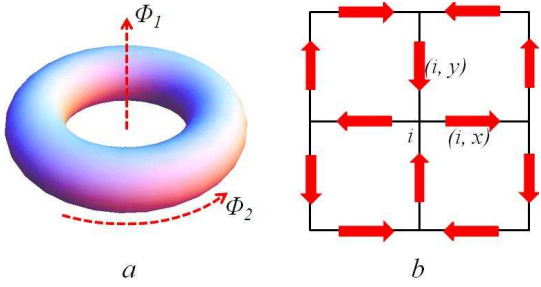}
\caption{($a$), the topological degeneracy can be counted as the
number of inequivalent commuting gauge fluxes $\Phi_1$ and
$\Phi_2$ through both cycles of the torus. ($b$), The ring
exchange terms Eq.~\ref{qring} are ring product of gauge field on
four links of each square, and in the ring exchange terms the
links are connected in the sequence of the arrows around each
plaquette.} \label{latticesign}
\end{figure}

\noindent{\it -- Type C, with G = $Q_8$:} Now let us move on to
the type $C$ phase, whose ground state is characterized by three
perpendicular nematic directors. This phase
can be obtained from  a system of spin-1 $SU(2)$ quantum spins
$\hat{S}^a_i$ possessing both two-spin and four-spin interactions
\cite{nitheory1,nitheory2}. For a spin-1 system it is often
convenient \cite{balentstrebst} to introduce SU(3) Schwinger
bosons $\vec{b}_i$. When $\langle \vec{b}^\ast \rangle $ is
parallel with $\langle \vec{b} \rangle $, the SO(3) spin symmetry
is broken, while there is no spin polarization on any site, thus
the system only has nematic order. Since $\langle \vec{b}^\ast
\rangle
\parallel \langle \vec{b} \rangle $,
the Schwinger boson $\vec{b}$ can be rewritten as $\langle
\vec{b}_i \rangle = e^{i\theta} \vec{N}_i$. $\vec{N}_i$ is in fact
a nematic director, because the transformation $\vec{N}
\rightarrow -\vec{N}$ can be cancelled by the transformation
$\theta \rightarrow \theta + \pi$, which is part of the U(1) gauge
symmetry associated with the Schwinger boson $\vec{b}$. The
vectors $\vec{N}_i$ are precisely the nematic directors in
Fig.~\ref{nematicfig}$C$.

Experimentally it was observed that the triangular lattice spin-1
material $\mathrm{NiGa_2S_4}$ has no global spin order with
time-reversal symmetry breaking at low temperature, but it still
has gapless excitations with linear dispersion \cite{nigas2}. It
has been proposed\cite{nitheory1,nitheory2} that the candidate
ground state of this system is characterized by a `tetrad' of
three perpendicular nematic directors $N^a_i$ on three different
sublattices denoted by $i=1,2,3$.
This proposed ground state thus has exactly the same GSM as that
of Fig.~\ref{nematicfig}$C$. The physical order parameter of this
state is the quadrupolar spin order parameter\footnote{Different
such quadrupolar spin order can exist on the three sublattices of
the triangular lattice}: \beqn Q^{ab}_i \sim N^a_iN^b_i -
\frac{1}{3}(\vec{N}_i)^2 \sim \langle \hat{S}^a_i\hat{S}^b_i -
\frac{2}{3} \rangle. \label{QuadrupolarOrder}
 \eeqn
Here $\hat{S}^a_i$ is the spin-1 operator on sublattices $i =
1,2,3$. This equation defines $N^a_i$, and it is precisely the
nematic director $N^a_i$ in the previous paragraph introduced
through Schwingber bosons \cite{balentstrebst}. This GSM is
equivalent to that of the biaxial nematic order \footnote{The
non-Abelian vortices of biaxial nematic order were also discussed
in the context of liquid crystals \cite{volovikquaternion}.} of a
liquid crystal \cite{biaxial1,biaxial2}. Similar ``triatic"
nematic spin order was also found in numerical work on SU(2)
spin-1/2 models with both two-spin and four-spin interactions on
the triangular lattice \cite{momoi2006}.

It is still most convenient to describe this phase with the SU(2)
rotor variable $\mathcal{Z}$, but now $\mathcal{Z}$ is coupled to
a discrete non-Abelian gauge field taking values in the
non-Abelian Quaternion group $Q_8$, \beqn Q_8 = \{ \pm 1, \ \pm i
\sigma^x, \ \pm i\sigma^y, \ \pm i\sigma^z\}. \label{gq} \eeqn
Again, the gauge field acts on the rotor field $\mathcal{Z}$ by
right multiplication. As a consequence of the action of this gauge
group, $\vec{N}_i$ in Eq.~\ref{rotor} become headless nematic
directors.

Now we will describe this gauge theory based on the non-Abelian
quaterion group $Q_8$ in more detail. Following the general
construction in Ref.~\cite{kitaev1997}, we define an 8-dimensional
Hilbert space $\mathcal{H}$ on each link $({i,\mu})$ of the
lattice, whose basis elements we denote by $|g_{i,\mu} \rangle$.
Here $i$ denotes a lattice site, $\mu=\hat{x},\hat{y}$ a unit
vector in a (positive) lattice direction, and $g_{i,\mu}$ denotes
any of the eight elements of the group  $Q_8$. Now we define, for
any group element $h\in Q_8$, and on every link ${i,\pm\mu}$ of
the lattice, operators $T^h_{i,\pm\mu}$ and $Q^h_{i,\pm\mu}$ with
the following action on the basis vector $| g_{i,\mu} \rangle$
residing on that link: \beqn && T^h_{i,+\mu} |g_{i,\mu} \rangle =
\delta_{h,g_{i,\mu}} |g_{i,\mu}\rangle, \ \ T^{h}_{i+\mu,-\mu}
|g_{i,\mu}\rangle = \delta_{h^{-1},g_{i,\mu}} |g_{i,\mu}\rangle,
\cr\cr && Q^h_{i, + \mu} |g_{i,\mu}\rangle = |hg_{i,\mu}\rangle, \
\ Q^h_{i+\mu,-\mu} |g_{i,\mu}\rangle = |g_{i,\mu}h^{-1}\rangle,
\cr\cr && (T^h_{i,\mu})^\dagger = T^h_{i,\mu} = T^{h^{-1}}_{i+\mu,
-\mu}, \cr \cr && (Q^{h}_{i,\mu})^{-1} = (Q^{h}_{i,\mu})^{\dagger}
= Q^{(h^{-1})}_{i,\mu}. \eeqn $T^h_{i,\mu}$ and $Q^f_{i,\mu}$ turn
out to satisfy the following algebra: \beqn &&
Q^f_{i,\mu}T^h_{i,\mu} Q^{(f^{-1})}_{i,\mu} = T^{fh}_{i,\mu},
\cr\cr && Q^f_{i+\mu, -\mu}T^h_{i,\mu} Q^{(f^{-1})}_{i+\mu, -\mu}
= T^{hf^{-1}}_{i,\mu}. \eeqn The dynamics of the discrete gauge
field is given by the following `ring exchange' term
\footnote{Note that by definition $h_{i,\mu} = h_{i+\mu, -\mu}$.}:
\beqn && H^{q}_{\mathrm{ring}} = \sum_{h} - K \ T^{h_{i + \hat{x},
\hat{y}}}_{i+\hat{x}, \hat{y}}T^{h_{i + \hat{x} + \hat{y}, -
\hat{x}}}_{i + \hat{x} + \hat{y}, - \hat{x}}T^{h_{i+\hat{y}, -
\hat{y}}}_{i+\hat{y}, - \hat{y}} T^{h_{i,\hat{x}}}_{i,\hat{x}}
\cr\cr &\times& \mathrm{tr}[G_{h_{i + \hat{x}, \hat{y}}}G_{h_{i +
\hat{x} + \hat{y}, - \hat{x}}}G_{h_{i+\hat{y}, -
\hat{y}}}G_{h_{i,\hat{x}}}] + H.c. \label{qring}\eeqn $G_h$ is the
two dimensional representation Eq.~\ref{gq} of the group element
$h\in Q_8$, and $\sum_h$ denotes summation over all group elements
$h_{i,\mu}\in Q_8$ on each link.

The direction of $\mu$ in the ring exchange term on each plaquette
follows the arrows in Fig.~\ref{latticesign}. Here we always
assume $K > 0$, which favors the gauge flux through each plaquette
to be 1.

The SU(2) rotor field $\mathcal{Z}_i$ is defined on the vertices
$i$ of the square lattice.
Right- and left- multiplication of $\mathcal{Z}$ by SU(2)
transformations $\mathrm{SU(2)_{right}}$ and
$\mathrm{SU(2)_{left}}$ is generated by the operators $J^a_{R}$
and $J^a_{L}$, satisfying the commutation relations (see also
Ref.~\onlinecite{Hermele2007})
\beqn [J^a_{R,L}, J^b_{R,L}] &=& i \epsilon_{abc} J^c_{R,L}, \ \
[J^a_{R}, J^b_{L}] = 0. \label{SU2RL} \eeqn In particular,
$J^a_{L}$ and $J^a_{R}$  act as follows: \beqn
e^{i\vec{\theta}\cdot \vec{J}_R} \mathcal{Z} e^{-
i\vec{\theta}\cdot \vec{J}_R} = \mathcal{Z} e^{
-i\frac{\vec{\theta}\cdot\vec{\sigma}}{2}}; \ \
e^{i\vec{\theta}\cdot \vec{J}_L} \mathcal{Z} e^{-
i\vec{\theta}\cdot \vec{J}_L } = e^{ i
\frac{\vec{\theta}\cdot\vec{\sigma}}{2}} \mathcal{Z}.  \eeqn The
quaternion gauge group is a subgroup of the
$\mathrm{SU(2)_{right}}$ transformation.

The full Hamiltonian with both, rotor and gauge fields reads
\beqn H \ \ \  &=& H_{rot}
 -\sum_{i,\mu, h_{i,\mu}} t \
\mathrm{tr}[\mathcal{Z}_i T^{h_{i,\mu}}_{i,\mu} G_{h_{i,\mu}}
\mathcal{Z}^\dagger_{i+\mu}] + H^{q}_{\mathrm{ring}} \cr\cr
H_{rot}&=&   \sum_{i} \sum_a \frac{U_R}{2} J^{a 2}_{R, i} +
\frac{U_L}{2} J^{a 2}_{L, i} \ \label{hamiltonian} \eeqn This
Hamiltonian is subject to the following quaternion gauge group
constraint: \beqn && e^{i \pi J^{a}_{R,i}} = Q^{i\sigma^a}_{i,
+\hat{x}} Q^{i\sigma^a}_{i, - \hat{x}} Q^{i\sigma^a}_{i, +
\hat{y}} Q^{i\sigma^a}_{i, - \hat{y}}. \label{VertexConstraint}
\eeqn $a = x, y, z$. The unitary operator $T^{h_{i,\mu}}_{i,\mu}
G_{h_{i,\mu}}$ appearing in Eq. \ref{hamiltonian} is the analogue
of the conventional term $e^{i\vec{A}_{i,\mu}\cdot \vec{\sigma}}$
where $\vec{A}_{i,\mu}$ is the gauge potential. The quaternion
group gauge constraint Eq.~\ref{VertexConstraint} generates the
following gauge transformations on both $\mathcal{Z}$ and
$T^{h_{i,\mu}}_{i,\mu}$: \beqn T^{h_{i,\mu}}_{i,\mu} \rightarrow
T^{f_{i}h_{i,\mu}f^{-1}_{i+\mu}}_{i,\mu}, \ \ \mathcal{Z}_i
\rightarrow \mathcal{Z}_i G_{f_{i}}, \eeqn where $f_i \in Q_8$.
The Hamiltonian Eq.~\ref{hamiltonian} is invariant under this
gauge transformation. We have formulated this model on the square
lattice, but generalizations to other lattices are
straightforward. Again, the quantum phase transition between the
ordered phase and quaternion liquid phase belongs to the 3D O(4)
universality class because the $Q_8$ gauge field is always gapped.

When $U_L, U_R \gg t$, the SU(2) rotor field $\mathcal{Z}_i$ is
gapped out, and the system is described by the pure quaternion
group gauge theory Eq.~\ref{qring}, plus the gauge constraints. In
the spin Hamiltonian, the rotor field $\mathcal{Z}_i$ can be
gapped by turning on the following term on the spin Hamiltonian
considered in Ref.~\cite{nitheory1,nitheory2}: \beqn H^\prime =
\sum_{ \ll i,j \gg }J^\prime \hat{Q}_i \cdot \hat{Q}_j, \ \ \
J^\prime >  0. \label{afq} \eeqn where $\hat{Q}_i$ is the
five-component quadrupole order parameter introduced in
Ref.~\cite{nitheory1,nitheory2,balentstrebst}. Eq.~\ref{afq} is an
antiferro-quadrupole interaction between the 2nd neighbor sites on
the triangular lattice. This term energetically disfavors the
system to form a three sublattice tetrad nematic order, and we
propose that it will drive the system into the phase described by
the pure quaternion nonabelian gauge theory. This gapped phase is
a realization of the {\it non-Abelian toric code} phase built on a
finite group $G$, proposed by Kitaev \cite{kitaev1997}. In the
present case $G=$ $Q_8$. Due to the non-Abelian nature of the
group $Q_8$, this gauge theory is known to possess a rich set of
gapped excitations exhibiting non-Abelian statistics,  which are
characterized by the representations of the so-called Drinfeld
double \cite{DrinfeldDouble1,DrinfeldDouble2,DijkgraafWitten} of
the group $Q_8$. These  excitations are the following:

(i)  {\it magnetic} excitations are located at the centers of the
plaquettes of the lattice (see Fig. 3), and are characterized by
the  product of group elements around a plaquette. Since the
product can be taken over different closed loops enclosing the
same ``magnetic flux'', a magnetic excitation is not characterized
by a group element $g$ itself, but by its conjugacy class ${\cal
C}_g=$ $\{ h^{-1} g h: \ h \in G\}$.

(ii) {\it electric} charges are located at the vertices of the
lattice (see fig. 3).  An electric charge represents a violation
of the vertex constraint of Eq. \ref{VertexConstraint} and
corresponds to an irreducible representation $\alpha$ of the group
$G$. Transporting an electric charge $\alpha$ around a magnetic
flux ${\cal C}_g$ along a closed path yields the representation
matrix $D^{(\alpha)}(g)$ of the group element $g$.

(iii) the most general excitation contains both, magnetic and
electric charges (often called a ``dyon''), and is represented by
a pair $({\cal C}_g, a)$ as follows: when there is no magnetic
charge, ${\cal C}_g=$ ${\cal C}_{g=1}$, then  $a=\alpha$ is an
electric charge, i.e. a representation of the group $G$. However
when the magnetic charge associated with a ``dyon'' is not
vanishing, i.e. when ${\cal C}_g \not =$ ${\cal C}_{g=1}$,
 its electric charge $a$ is an irreducible  representation
 $a={\hat \alpha}$ of the {\it Normalizer}
$N(g) =$ $\{ h \in G: h g= g h \}$  of $g$  (consisting of all
those group elements commuting with $g$), which is in general not
the entire group $G$, but only a subgroup thereof. -- Let us count
the total number of excitations for the Drinfeld double of the
quarternion group $Q_8$.  We use the following facts:  there  are
5 conjugacy classes
 $\{+1  \} , \{ -1 \}, \{  \pm i \sigma^a \}$
where $a=x,y,z$;  the number of irreducible representations  of
any  finite group equals the number of conjugacy classes; the
centralizer of any of the three conjugacy classes $\{  \pm i
\sigma^a \}$ is the Abelian cyclic group $Z_4$ of four elements
generated by $i\sigma^a$. Thus, there are 5 excitations of the
form $({\cal C}_1, \alpha )$, 5 of the form $({\cal C}_{-1},
\alpha )$, and 4 of the form $({\cal C}_{i\sigma^a}, {\hat
\alpha})$, for each $a=x,y,z$, where ${\hat \alpha}$ labels the
representations of $Z_4$. This amounts to a total of $5+5+3 \times
4=$ $22$ excitations. Since for a general 2D topological field
theory the ground state degeneracy on the torus equals the number
of topological excitations (``particles''), this degeneracy is 22
in the present case. The fusion rules for these 22 particles can
be obtained from the modular S-matrix (through the Verlinde
Formula) which, in turn, can be obtained in the standard
manner\cite{DrinfeldDouble2} from the Drinfeld double
construction.

In general,  for topological phases
which are Drinfeld doubles of a finite group, the number of
topological sectors on the torus corresponds precisely to the
number of commuting pairs of gauge inequivalent magnetic fluxes
through the two cycles of the torus (Fig.~\ref{latticesign}$a$).
(The two fluxes need to commute in order to keep the system in its
ground state.)

It is important to note that the statistics of the excitations is
only well-defined in the disordered liquid phase (described by
pure $Q_8$ gauge theory). The ordered phase with a
$\mathcal{Z}$-condensate has gapless Goldstone modes, which make
adiabatic braiding operations impossible. Another important
difference between the ordered and the disordered phase is that
these non-Abelian defects have logarithmic divergent energy in the
ordered phase, while in the disordered phase they all have finite
energy.

{\it Summary:} In this work we studied a fully gapped topological
spin liquid state with non-Abelian excitations. Despite its
complicated effective model description, we propose that such
state can be realized by disordering a rather simple spin order of
a spin-1 quantum SU(2) magnet.


\begin{acknowledgments}

{\it Acknowledgement:} The authors are grateful to Zhenghan Wang
for illuminating discussions on quantum doubles.
This work was supported, in part, by the NSF under Grant No. DMR-
0706140 (A.W.W.L.).

\end{acknowledgments}

\bibliography{nematicliquid}

\end{document}